\documentclass[twocolumn,showpacs,preprintnumbers,amsmath,amssymb,superscriptaddress,revsymb,prb]{revtex4}

\usepackage[dvips]{graphicx}
\usepackage{mathptmx}
\usepackage{dcolumn}

\newcommand{\etal}{\emph{et al.}}

\newcolumntype{.}{D{.}{\cdot}{3.10}}

\begin{document}

\title{Thermodynamically stable lithium silicides and germanides from density-functional theory calculations}

\author{Andrew J. Morris\footnote{Email: ajm255@cam.ac.uk}}
\affiliation{Theory of Condensed Matter Group, Cavendish Laboratory, University of Cambridge, J. J. Thomson Avenue, Cambridge CB3 0HE, United Kingdom}
\author{C. P. Grey} \affiliation{Department of Chemistry, University of Cambridge, Lensfield Road, Cambridge CB2 1EW, United Kingdom}
\author{Chris J. Pickard}
\affiliation{Department of Physics and Astronomy, University College London, Gower St, London WC1E 6BT, United Kingdom}

\date{\today{}}

\begin{abstract} 
High-throughput density-functional-theory (DFT) calculations have been performed on the Li-Si and Li-Ge systems.
Lithiated Si and Ge, including their metastable phases, play an important technological r\^ole as Li-ion battery (LIB) anodes.
The calculations comprise structural optimisations on crystal structures obtained by swapping atomic species to Li-Si and Li-Ge from the X-Y structures in the International Crystal Structure Database, where X$=$\{Li,Na,K,Rb,Cs\} and Y$=$\{Si,Ge,Sn,Pb\}.
To complement this at various Li-Si and Li-Ge stoichiometries, \emph{ab initio} random structure searching (AIRSS) was also performed. 
Between the ground-state stoichiometries, including the recently found Li$_{17}$Si$_{4}$ phase, the average voltages were calculated, indicating that germanium may be a safer alternative to silicon anodes in LIB, due to its higher lithium insertion voltage.
Calculations predict high-density Li$_1$Si$_1$ and Li$_1$Ge$_1$ $P4/mmm$ layered phases which become the ground state above 2.5\, and 5\,GPa respectively and reveal silicon and germanium's propensity to form dumbbells in the Li$_x$Si, $x=2.33-3.25$ stoichiometry range.
DFT predicts the stability of the Li$_{11}$Ge$_6$ $Cmmm$, Li$_{12}$Ge$_7$ $Pnma$ and Li$_7$Ge$_3$ $P32_12$ phases and several new Li-Ge compounds, with stoichiometries Li$_5$Ge$_2$, Li$_{13}$Ge$_5$, Li$_8$Ge$_3$ and Li$_{13}$Ge$_4$.

\end{abstract}

\maketitle

 \section{Introduction}

Lithium-ion batteries (LIBs) are the secondary (rechargeable) battery
of choice for portable electronic devices due to their high specific
energy (energy per unit weight) and energy density (energy per unit
volume).
LIBs have the highest capacity of all the commercially available
battery technologies and are now being deployed in hybrid and
all-electric vehicles.\cite{chu:nature:2012}
There is substantial interest in enhancing the capacity of LIBs,
driven by the economic and environmental advantages of increasing the
range of electric vehicles, and enabling longer-life portable
electronic devices.

Lithium intercalated graphite is the standard LIB negative electrode
material due to its good rate capability and cyclability, but demand
for even higher performance LIBs has motivated the investigation of
other materials.
Silicon is an attractive alternative since it has ten times the
gravimetric and volumetric capacity of graphite (calculated from the
initial mass and volume of silicon) but, unlike graphite, silicon
undergoes structural changes on
lithiation.\cite{lai:JES:1976,wen:JSSC:1981,weydanz:JPS:1999}
The negative electrode may be studied using a half-cell containing
lithium and silicon.
The term ``anode'' applies to the negative electrode during LIB
discharge only, so to avoid confusion we refer to \emph{lithiation}
and \emph{delithiation} of the silicon half-cell which corresponds to
charging and discharging the LIB respectively.
The first lithiation of the cell at room temperature involves the
conversion of crystalline silicon ($c$-Si) into an amorphous lithium
silicide phase ($a$-Li$_y$Si).\cite{amorphous_LiSi}
The onset of amorphization depends on the lithiation rate and has been
measured at $y \approx 0.3$ in micron-sized (325 mesh) silicon
clusters after irreversible SEI (solid-electrolyte interphase)
formation has been taken into
account.\cite{key:JACS:2009,li:JES:2007}
Below a discharge voltage of 50\,mV the $a$-Li$_y$Si crystallizes to
form a metastable Li$_{15}$Si$_{4}$ phase which may become non-stoichiometric,
Li$_{15\pm\delta}$Si$_{4}$, at deep discharge.\cite{key:JACS:2009}
However, at temperatures above 100$^{\mathrm o}$C it is possible to form the most lithiated crystalline phase, Li$_{21}$Si$_5$, electrochemically.\cite{kwon:AE:2010}
Full lithiation of silicon leads to a drastic volume expansion of
270--280\%,\cite{obrovac:JES:2007} which generates considerable
mechanical stress.
Hysteresis in the capacity/voltage profile occurs due to a combination
of mechanical stress and different reactions taking different
structural pathways on lithiation and delithiation.
The microscopic mechanisms underlying these phenomena are still not
entirely clear.\cite{obrovac:JES:2007}
$a$-Li$_y$Si has been studied \emph{in situ} using nuclear magnetic
resonance (NMR),\cite{key:JACS:2009} X-ray diffraction (XRD)
\cite{Hatchard:JES:2004,Obrovac:ESSL:2004} and electron energy loss
spectroscopy (EELS). \cite{kang:IC:2009}
These studies along with \emph{ex situ} NMR and PDF (pair-distribution
function) studies of XRD data suggest that silicon forms small
clusters and isolated atoms during lithiation.
The clusters that form only break apart into isolated silicon atoms at
the end of the lithiation process (below 80\,mV). \cite{key:JACS:2011}

Many of the disordered structures that form during lithiation can be approximated by the Li-Si ground-state and metastable crystalline phases.
For instance, the crystalline phases have been used as a first step in understanding charge transfer \cite{chevrier:JAC:2010}  and average lithiation voltages.\cite{chevrier:CJP:2009}
To gain insight into the possible types of silicon clusters present and
their environments in $a$-Li$_y$Si, various crystalline phases have been investigated and
categorized using NMR \cite{key:JACS:2009,koester:AC:2011} and
\emph{ab initio} theoretical techniques.
\cite{chevrier:JES:2010,chevrier:JES:2009,chevrier:JAC:2010,Zeilinger:CoM:2013:Li17Si4}
These $c$-Li-Si phases have previously been well categorized using density-functional theory (DFT),\cite{chevrier:JES:2010,chevrier:JES:2009,chevrier:JAC:2010} however new insights into the most lithiated phases and the ability to synthesize Li$_1$Si$_1$ through ball milling have suggested that the system is far from fully understood.
The most recent phase diagram of the Li-Si system shows, in ascending
lithium content order, $c$-Si, Li$_1$Si$_1$, Li$_{12}$Si$_{7}$, Li$_{7}$Si$_{3}$, Li$_{13}$Si$_{4}$,
Li$_{15}$Si$_{4}$,  Li$_{22}$Si$_{5}$, and $\beta$-Li.\cite{okamoto:JPED:2009}
Additionally investigations by Zeilinger and coworkers have presented a high-temperature Li$_{4.11}$Si phase \cite{Zeilinger:CoM:2013:Li17Si4,Zeilnger:CoM:2013:Li4.11Si} and suggested Li$_{17}$Si$_{4}$ as the correct stoichiometry of Li$_{21}$Si$_{5}$/Li$_{22}$Si$_{5}$.

Germanium is another choice of anode for LIB with a theoretical capacity of 1568\,mAh\,g$^{-1}$ some 5 times greater than carbon.
Its lithium diffusivity at room temperature is 400 times greater than silicon, \cite{park:CSR:2010} however it is  scarcer and consequently more expensive.
About the Li-Ge phase diagram, much less is known.
In increasing order of lithium content, the following stable phases have all been proposed: Li$_7$Ge$_{12}$,\cite{gruttner:ACA:1981} Li$_1$Ge$_1$,\cite{menges:ZN:1969,evers:ACIE:1987} Li$_{12}$Ge$_7$,\cite{gruttner:ACA:1981} Li$_{11}$Ge$_6$,\cite{frank:ZN:1975} Li$_9$Ge$_4$,\cite{hopf:ZN:1970,jain:JPPC:2013,yoon:ESSL:2008} Li$_7$Ge$_3$,\cite{gruttner:ACA:1981,jain:JPPC:2013} Li$_7$Ge$_2$,\cite{hopf:ZN:1972,yoon:ESSL:2008} Li$_{15}$Ge$_{4}$,\cite{gladyshevskii:SPC:1961,johnson:AC:1965,jain:JPPC:2013,yoon:ESSL:2008}  Li$_{17}$Ge$_4$\cite{goward:JAC:2001} and Li$_{22}$Ge$_5$.\cite{gladyshevskii:SPC:1964,yoon:ESSL:2008,jain:JPPC:2013}
More crystalline phases occur during electrochemical lithiation of germanium than silicon; XRD and HRTEM measurements show that during lithiation of germanium at room temperature, the Li-Ge system progressed through Li$_9$Ge$_4$, Li$_7$Ge$_2$ and a mixture of Li$_{15}$Ge$_4$ and Li$_{22}$Ge$_5$. \cite{yoon:ESSL:2008}

In this article we use atomic species swapping along with random structure searching techniques, described in Sec.\,\ref{methods}, to predict ground state and metastable crystal structures of the Li-Si and and Li-Ge systems.
In Sec.\,\ref{voltages} our approach to calculating average voltages is discussed and Secs\,\ref{LiSi} and \,\ref{LiGe} describe the DFT-predicted phases of Li-Si and Li-Ge respectively.
In the Li-Si system we predict a high-density Li$_1$Si$_1$ phase with $P4/mmm$ symmetry and discuss the tendency for silicon to form dumbbells within the lithium silicides.
We then turn our attention to Li-Ge which has not been analyzed using these computational search methods method before and predict the new structures, Li$_5$Ge$_2$, Li$_{13}$Ge$_5$, Li$_8$Ge$_3$ and Li$_{13}$Ge$_4$. 
The average voltages for the Li-Si and Li-Ge systems are presented including the Li$_{17}$Si$_4$  and Li$_1$Si$_1$ phases. 
The conclusions of the simulations are given in Sec.\,\ref{Sect:Conclusions}.

\section{Methods}
\label{methods}

\emph{Ab initio} random structure searching (AIRSS) has been successful in predicting the ground-state structures of high-pressure phases of matter.\cite{pickard:PRL:2006,Pickard_AIRSS}
More recently is has also been applied to the Li-P system \cite{ivanov:ACIE:2012} and defects in technologically relevant ceramics, \cite{mulroue:PRB:2011,mulroue:JNM:2013} semiconductors \cite{morris:PRB:2008,morris:PRB:2009} and LIBs. \cite{morris:PRB:2011, morris:PRB:2013}
Since in an AIRSS calculation each random starting configuration is independent from another,  the search algorithm is trivially parallelisable, making high-throughput computation straightforward.
AIRSS searches were performed for stoichiometries Li$_x$Si$_y$ and Li$_x$Ge$_y$ where $x,y=1-8$.

Relaxations were performed using the stoichiometric crystal structures of Li-Si, Li-Ge, Li-Sn, Li-Pb, Na-Si, Na-Ge, Na-Sn, Na-Pb,
K-Si, K-Ge, K-Sn and K-Pb.
First, the structures were obtained from the International Crystallographic Structure Database (ICSD).
Second, for each structure the anions were replaced with Li and the cations replaced with \{Si,Ge\}.
The structures were relaxed to local-energy minima using forces and stresses calculated by DFT.

Calculations were performed using the plane wave \textsc{castep} DFT code.\cite{CASTEP:ZK:2004}
The PBE (Perdew-Burke-Ernzerhof) exchange-correlation
functional was used with Vanderbilt ``ultrasoft'' pseudopotentials.
The Li-Si system required a basis set containing plane waves with energies of up to 400\,eV and
a Monkhorst-Pack (MP) grid corresponding to a Brillouin zone (BZ) sampling grid finer
than $2\pi \times 0.05\,\AA^{-1}$.
The Li-Ge system required a 600\,eV planewave cutoff with harder pseudopotentials
and a BZ sampling finer than $2\pi\times0.03\,\AA^{-1}$.

We define the formation energy per atom of a compound Li$_m$X$_n$, where $X=\{$Si,Ge$\}$ as,
\begin{equation}
E_f/A = \frac{E\left ( {\rm Li}_{n}{\rm X}_{m} \right ) - n\mu_{\rm
  Li} - m\mu_{\rm X}}{n+m},
\end{equation}
where $E\left ( {\rm Li}_{n}{\rm X}_{m} \right )$ is the total DFT
energy of a given structure, Li$_n$X$_m$ and $\mu_{\rm Li}$ and $\mu_{\rm X}$ are the chemical potentials of  atomic species Li and X in their ground state elemental structure.
To compare the stabilities of different stoichiometries we plot the
formation energy per atom, $E_f/A$ versus the fractional concentration
of lithium in a compound where,
\begin{equation}
C_{\rm Li} = \frac{n}{n+m},
\end{equation}
and, as above, $n$ and $m$ are the number of atoms of Li and X in a compound respectively.
Drawing a convex hull from $(C_{\rm Li}, E_f/A)=(0,0)$ to $(1,0)$, that is, between the chemical potentials, reveals the stable zero Kelvin structures at the vertices of the tie-lines.

\section{Voltages}
\label{voltages}

We calculate average voltages in an LIB anode using DFT total energies by assuming that all the displaced charge is due to Li and that the reaction proceeds sequentially through the phases on the tie-lines of the convex hull, \emph{i.e.} it is a succession of two-phase reactions.\cite{aydinol:PRB:1997}
The voltage is given by, 
\begin{equation}
V = -\frac{\Delta G}{\Delta x},
\end{equation}
where the Gibbs free energy change, $\Delta G$ is in eV and $\Delta x$
is the change in the number of lithium atoms per silicon atoms in the 2 phases.
The Gibbs free energy is composed of a number of terms,
\begin{equation}
\Delta G = \Delta E + P\Delta V -T \Delta S, 
\end{equation}
where $ \Delta E$ is the total electronic energy, and $P$, $\Delta
V$, $T$ and $\Delta S$ are the pressure, change in volume,
thermodynamic temperature and change in entropy respectively.
Due to the difficulty in calculating $\Delta G$ we make the approximation, previously applied to the Li-Si system, that $\Delta G \approx \Delta E$ since $\Delta E$ is of the order of a few electron volts, $P\Delta V\approx 10^{-5}$\,eV and $T\Delta S \approx 0.06$\ eV at $425\,^{\mathrm o}C$. \cite{chevrier:CJP:2009,courtney:PRB:1998,tipton:PRB:2013}
%

\section{Results - Lithium Silicide}
\label{LiSi}

\begin{center}
\begin{figure*}
\includegraphics*[width=150mm]{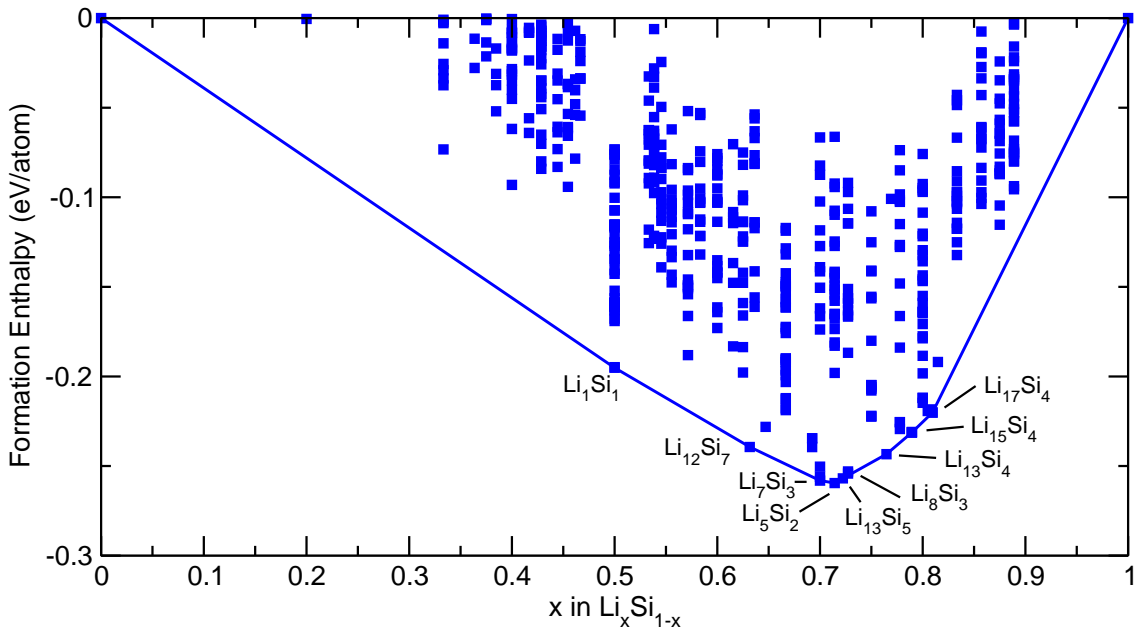}
\caption[]{(Color online) Formation enthalpy per atom versus the fractional lithium concentration of a Li-Si compound (blue boxes).  The tie-line (blue line) shows the convex-hull obtained by joining together the globally stable structures predicted by DFT. Between the species swapping technique and AIRSS we recover the known stable phases, Li$_1$Si$_1$, Li$_{12}$Si$_7$, Li$_7$Si$_3$, Li$_{13}$Si$_4$, Li$_{15}$Si$_4$ and Li$_{17}$Si$_4$. The searches also find the Li$_5$Si$_2$ phase predicted by Tipton \emph{et al.} \cite{tipton:PRB:2013} and  low-lying metastable phases with stoichiometries Li$_8$Si$_3$ and Li$_{13}$Si$_5$ are also predicted.}
\label{Fig:LiSihull}
\end{figure*}
\end{center}

We find on the convex hull, shown Fig.\,\ref{Fig:LiSihull}, in increasing lithium content order;
diamond-structure $Fd\overline{3}m$ $c$-Si;
the $I4_1/a$ Li$_1$Si$_1$ phase\cite{stearns:JSSC:2003,kubota:JAC:2008} which has recently been synthesized at ambient pressure \cite{tang:JES:2013} and is discussed in Sec.\,\ref{Sect:Li1Si1}; 
the $Pnma$ Li$_{12}$Si$_7$ phase \cite{nesper:CB:1986} which contains
silicon 5-membered rings and 4-membered stars, which have been studied using NMR;\cite{koester:AC:2011}
the Li$_{7}$Si$_{3}$ and Li$_{5}$Si$_{2}$ phases with $P3_212$ and
$R\overline{3}m$ symmetries respectively discussed further in Sec.\,\ref{Sect:Li5Si2};
the $Pbam$ Li$_{13}$Si$_{4}$ phase;
the metastable $I4\overline{3}d$  Li$_{15}$Si$_{4}$; 
and the Li$_{17}$Si$_{4}$ $F\overline{4}3m$ symmetry phase discussed further in Sec.\,\ref{Sect:Li17Si4}.

For Li$_{15}$Si$_{4}$, Mulikan analysis yields a charge of  $0.15|e|$ and $-0.57|e|$ per Li and Si respectively; in agreement with Bader analysis that Li is a cation\cite{chevrier:JAC:2010} and contrary to the reports that Li is anionic.\cite{kubota:JAP:2007}

The average voltage was calculated between all adjacent pairs of stable Li-Si phases on the convex hull including both the Li$_{17}$Si$_4$ phase and Li$_1$Si$_1$ phase recently synthesized at ambient pressure. 
Voltages were obtained from the DFT total energies, as described in Sec.\,\ref{voltages} and referenced to lithium metal.
The potential composition curve is presented in Fig\,\ref{Fig:LiSi_voltages} as is in agreement with previous experimental and theoretical work. 

\subsection{Li$_{1}$Si$_1$ layered structures}
\label{Sect:Li1Si1}

We find a set of structures with Li$_1$Si$_1$ stoichiometry, listed in Table\,\ref{Table:Li1Si1}, all within $\sim0.1\,$eV/f.u. of the ground state.
The DFT ground state at 0\,GPa is a $I4_1/a$ phase comprising a 3-fold coordinated silicon network hosting lithium tetrahedra similar to the \{4Li,$V$\} \emph{Zintl defect} in silicon. \cite{morris:PRB:2013}
Recently the $I4_1/a$ phase has been synthesized via ball-milling and shown to be stable under ambient conditions.\cite{tang:JES:2013}
Mulikan analysis yields a charge of $0.34|e|$ for each Li and $-0.34|e|$ for each Si establishing Li as cationic contrary to a previous analysis.\cite{kubota:JAC:2008}

DFT predicts a novel $P4/mmm$ phase with a formation energy of 0.07\,eV p.f.u. at 0\,GPa.
It is a layered structure comprising a two-dimensional (non-tetrahedrally) four-fold coordinated silicon lattice with lithium intercalated between the silicon sheets.
Since the silicon is four-fold coordinated it gains less of the lithium's charge than in the $I4_1/a$  phase; Mulikan analysis shows lithium atoms donate $0.22|e|$ each.
Our calculations show the  Li$_1$Si$_1$ system undergoes a phase transition from the $I4_1/a$  to the $P4/mmm$ phase at 2.5\,GPa.

The $I4_1/amd$ phase is isostructural to its Li$_1$Ge$_1$ analogue.
Li$_1$Ge$_1$ $I4_1/amd$ is stable at high pressure,\cite{evers:ACIE:1987} however Li$_1$Si$_1$  $I4_1/amd$  is not globally stable over the pressure range we studied (between 0 and 10\,GPa).

\begin{center}
\begin{figure*}
\includegraphics*[width=140mm]{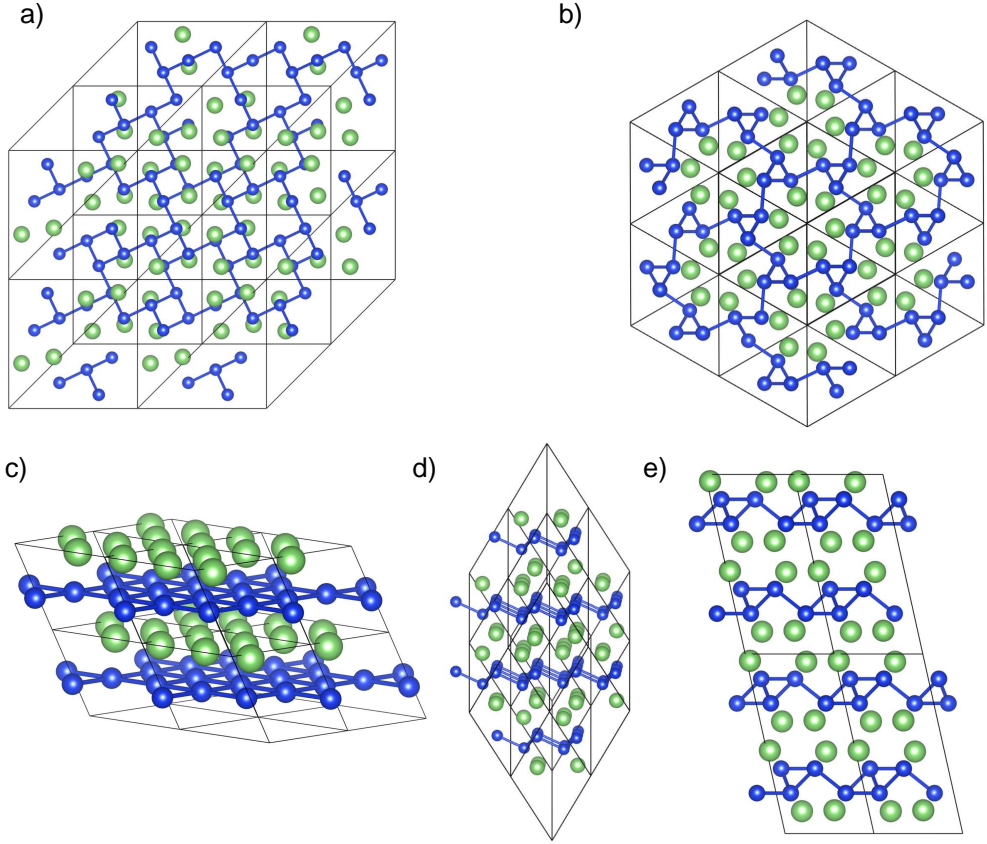}
\caption[]{(Color online) Low-energy Li$_1$Si$_1$ phases detailed in Table\,\ref{Table:Li1Si1} with a) $I4_1/a$, b) $R\overline{3}$, c) $P4/mmm$, d) $P\overline{3}m1$, e)$P\overline{1}$ symmetries. DFT predicts the $P4/mmm$ phase to be stable above 2.5\,GPa. }
\label{Fig:LiSi_montage}
\end{figure*}
\end{center}

The $P\overline{3}m1$ phase contains 6 membered rings of 3-fold coordinated silicon atoms in layered sheets, see Fig\,\ref{Fig:LiSi_montage}.
The silicon network is isostructural to silicene, a silicon analogue of graphene.
We calculate that silicene and a phase based on the $P4/mmm$ silicon network are 0.63 and 0.89\,eV p.f.u. respectively above the $Fd\overline{3}m$ ground state.
When lithiated both layered structures are only 0.07\,eV above the ($I4_1/a$) ground state.
Given the interest in silicene our layered compounds might provide an alternative route to layered silicon.

\begin{table*}
\caption{Low-energy Li$_1$Si$_1$ metastable phases. The structures are shown in Fig.\,\ref{Fig:LiSi_montage} with formation energy $E_f$ per formula unit relative to the energy of the ground states. We calculate that $P4/mmm$ is the most stable above 2.5\,GPa.}
\begin{tabular}{ @{} c  c c c}
\hline\hline
E$_f$ &  Symmetry & Volume   & Description \\
(eV/f.u.) & & (\AA$^3$/f.u.)  & \\
\hline
0.00    &   $I4_1/a$  &  31.3    &   Li tetrahedra in a 3-fold coordinated Si network \\
 0.05      &$R\overline{3}$  & 33.1    &      Distorted Li octahedra 3-fold coordinated Si network\\
    0.07   &   $P4/mmm$  &    27.8  &        Flat Si sheets comprising 4 membered rings \\
 0.07          &   $P\overline{1}$    &  31.7    &      Buckled Si sheets comprising 8 and 4 membered rings\\
 0.07   &      $P\overline{3}m1$  &       34.1   &     Li intercalated silicene  \\
  0.08     &   $P2/m$      &       28.1  &   Li$_1$Sn$_1$-like 2.39\,\AA\ dumbbells and isolated atoms \\
0.11 &  $I4_1/amd$    &  28.1   &        Isostructural with Li$_1$Ge$_1$  high-pressure phase\\
\hline\hline
\label{Table:Li1Si1} 
\end{tabular}
\end{table*}

\subsection{Li$_{7}$Si$_3$ and Li$_{5}$Si$_{2}$}
\label{Sect:Li5Si2}

Lithium's position in the crystal lattice can be difficult to establish due to its low XRD scattering factor.
Furthermore, Li$_{7}$Si$_{3}$ has partially occupied lattice sites\cite{vonSchnering:ZM:1980} making it difficult to model using DFT. 
Its structure may be represented as a supercell of  $R\overline{3}m$ Li$_5$Si$_2$ in
which lithium atoms have been removed from certain lattice sites. \cite{barvik:CJPB:1983}
By choosing different combinations of lithium sites in the supercell, models of Li$_{7}$Si$_{3}$ can be
produced with  $P\overline{3}m1$, $C2/m$ $Cm$ and $P3_212$ symmetries.
The latter, labeled ``\#2'' by Dahn \emph{et al.}, \cite{chevrier:JAC:2010} is found on the convex hull. 

It is unsurprising that at zero Kelvin, DFT also predicts that the $R\overline{3}m$ Li$_5$Si$_2$ phase to be stable since it contains entirely occupied lithium sites.
Tipton \emph{et al.} also found this phase to be stable using DFT. \cite{tipton:PRB:2013}
%

\subsection{Most lithiated phases}
\label{Sect:Li17Si4}

The most lithiated stable Li-Si phase has been the subject of debate. 
XRD measurements predict that Li$_{21}$Si$_5$ \cite{nesper:JSSC:1987} is stable at room temperature and Li$_{22}$Si$_5$ at 415$^o$C \cite{wen:JSSC:1981}.
Previous DFT calculations predict Li$_{21}$Si$_5$ to be the stabler phase, even after the inclusion of temperature dependence using the harmonic approximation.\cite{chevrier:JAC:2010}
The combined AIRSS/species-swapping technique predicts Li$_{21}$Si$_5$ and  Li$_{22}$Si$_5$ to be locally stable but above the convex hull.
The  Li$_{17}$Si$_4$ phase  is on the convex hull, and it has the same crystal structure as $F\overline{4}3m$ Li$_{17}$Pb$_4$, as discovered independently by Zeilinger \etal \cite{Zeilinger:CoM:2013:Li17Si4}
Zeilinger \etal also predict a Li$_{4.11}$Si high-temperature phase which they model using Li$_{16}$Si$_4$ and Li$_{16.5}$Si$_4$ phases.
We include Zeilinger \etal's models in Fig.\,\ref{Fig:LiSihull} although DFT predicts that they are not on the tie-line. \footnote{Since the Li$_{16.5}$Si$_4$ model contains partially occupied Li sites we extended the cell in the $a$ direction fully occupying four 8g and two 4c sites before optimizing the geometry using DFT.}

\subsection{Repeating units -- Silicon dumbbells}
\label{dumbells}

We also find a $R\overline{3}m$ Li$_8$Si$_3$ and a $P\overline{3}m1$  Li$_{13}$Si$_{4}$ phase close above the tie-line.
They belong to the set of  structures in the range Li$_7$Si$_3 \rightarrow$ Li$_{13}$Si$_5$ which all contain silicon dumbbells.
The dumbbells are aligned in parallel with various numbers of collinear lithium atoms between them forming one-dimensional repeating sequences, see Table\,\ref{table:columns}.
The 1D linear repeating chains are thus packed alongside each other realizing the three-dimensional structure.
For example, since Li$_7$Si$_3$ comprises (5$\times$Li + Si-Si) and (4$\times$Li + Si-Si) sequences in a ratio of 2:1, it is equivalent to the Li$_5$Si$_2$ phase with lithium vacancies.
Li$_5$Si$_{2}$ comprises sequences of atoms with the repeating unit (5$\times$Li + Si-Si) and the Li$_8$Si$_3$ is similar but with atoms in a (4$\times$Li + Si + 4$\times$Li + Si-Si) repeating unit; it is isostructural with a Li$_8$Pb$_3$ phase. \cite{cenzual:ZK:1990}
The Li$_{13}$Si$_{5}$  phase is isostructural with the Li$_{13}$Sn$_5$ phase \cite{frank:ZN:1975:Li3Sn5} and has two different repeating units (5$\times$Li + Si) and (4$\times$Li + Si-Si) in a ratio of 1:2.

Finally, in Li$_{13}$Si$_4$ the one-dimensional columnar structure does not exist but Si-Si dumbbells and Si isolated atoms remain in a ratio of 1:1.
At higher lithium concentrations, Li$_{15}$Si$_4$ forms, in which all silicon dumbbells are broken, and only isolated Si atoms remain.
The propensity for silicon dumbbells to form over a wide range of stoichiometries and total energies implies that silicon dumbbells form on lithiation of silicon in a LIB anode.

\begin{table}
\caption{\label{table:columns} Structures in the stoichiometry range Li$_x$Si, $x=2.33-2.60$ exhibit parallel silicon dumbbells situated in one-dimensional linear columns containing variable numbers silicon dumbbells and isolated lithium and silicon atoms. The columns are packed together realizing the three-dimensional structure. The repeating unit(s) in each column are represented between parentheses, and silicon dumbbells are indicated by Si-Si. For example, a structure comprising columns containing a repeating unit of 5 lithium atoms followed by a silicon dumbbell is represented as (5$\times$Li + Si-Si). }
\begin{tabular}{ @{}l c  }
\hline\hline
Stoichiometry & Constituent columns  \\
\hline
Li$_7$Si$_3$ & 2$\times$(5$\times$Li  + Si-Si) \&  \\
 & (4$\times$Li + Si-Si) \\
Li$_5$Si$_2$ & (5$\times$Li + Si-Si) \\
Li$_{13}$Si$_5$ & (5$\times$Li+Si) \& \\
& 2$\times$(4$\times$Li+Si-Si) \\
Li$_8$Si$_3$ & (4$\times$Li + Si + 4$\times$Li + Si-Si) \\
\hline\hline
\end{tabular}
\end{table}

\begin{figure}
\includegraphics*[width=90mm]{fig3.eps}
\begin{flushleft}
$^a$ Wen and Huggins. \cite{wen:JSSC:1981}\\
$^b$ Chevrier \etal\ \cite{chevrier:CJP:2009}
\end{flushleft}
\caption[]{(Color online) Potential-composition curves of stable structures found on the convex hull in Fig.\,\ref{Fig:LiSihull} (black line) compared to experiment at 415$^o$\,C (red dashed line) and previous DFT-GGA calculations (blue dot-dashed line). \cite{chevrier:CJP:2009}}
\label{Fig:LiSi_voltages}
\end{figure}

\section{Results - Lithium Germanide}
\label{LiGe}

\begin{center}
\begin{figure*}
\includegraphics*[width=150mm]{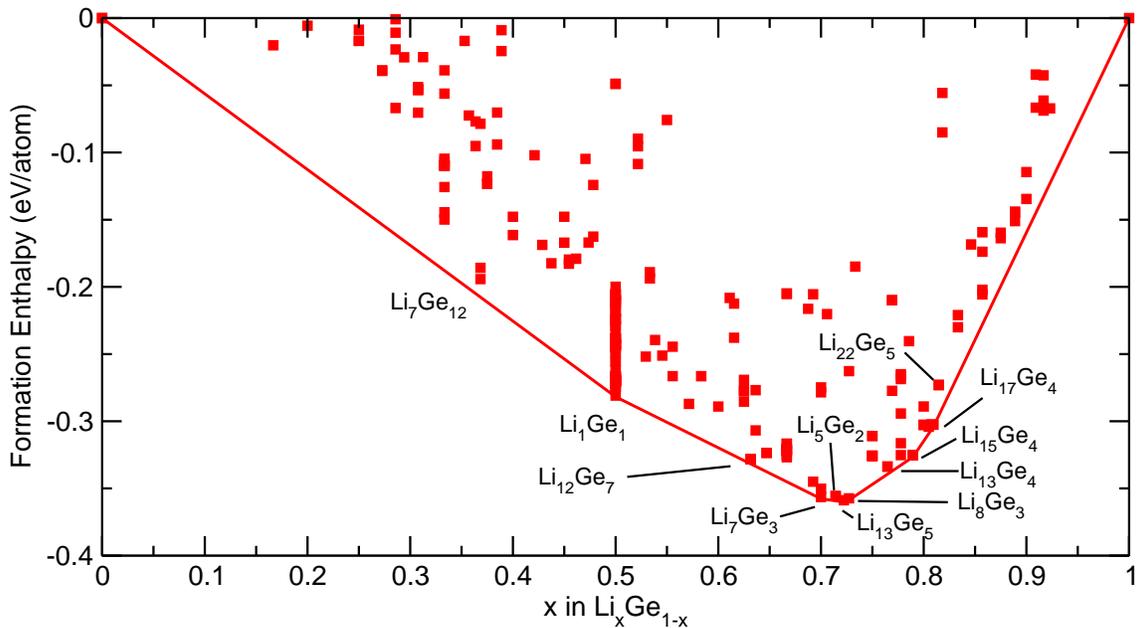}
\caption[]{(Color online) The Li-Ge binary composition diagram. The red squares indicate the formation enthalpy of a structure. The red line is the tie-line indicating the stable structures at 0\,K predicted by DFT.}
\label{Fig:LiGe_hull}
\end{figure*}
\end{center}

\begin{table*}
\label{table3}
\caption{Experimental and predicted phases of Li$_{\mathrm x}$Ge system.}
\begin{tabular}{ @{}l c  c c}
\hline\hline
Experimental Symmetry        & x         & Stoichiometry & Predicted Symmetry \\
\hline
$Fd\overline{3}m$$^{a}$         & 0.000 & Ge & $Fd\overline{3}m$\\
$Pmn2_1$$^{b}$, $P2/n$$^{c}$ & 0.580 &  Li$_{7}$Ge$_{12}$& $Pc$  \\
$I4_1/a$$^{d}$,$I4_1/amd^{*,e}$                        & 1.000 & Li$_1$Ge$_1$ & $I4_1/a$, $I4_1/amd^{*}$, $P4/mmm^{**}$\\
$Pnma$$^{f}$                         & 1.710 &  Li$_{12}$Ge$_7$ & $Pnma^{*}$\\
$Cmcm$$^{g}$                        &  1.83 &   Li$_{11}$Ge$_6$ & $Cmcm^{*}$ \\
$Cmcm$$^{h,i,j}$                  &  2.25 &   Li$_{9}$Ge$_4$ & $Cmcm^{*}$\\
$P32_12$$^{f}$, $R\overline{3}m$$^{i}$   &  2.33 & Li$_{7}$Ge$_3$ & $P32_12$, $P21/m^{*}$\\
                                              & 2.50 & Li$_{5}$Ge$_2$   &  $R\overline{3}m^*$ \\
                                              & 2.60 &  Li$_{13}$Ge$_5$  & $P\overline{3}m1$  \\
                            & 2.67 &   Li$_{8}$Ge$_3$ &   $R\overline{3}m$  \\
 $^{l}$ & 3.20 & Li$_{16}$Ge$_5$ & \\
                          & 3.25 &   Li$_{13}$Ge$_4$  &  $Pbam^*$\\
$Cmmm$$^{m,j}$           & 3.50 &  Li$_{7}$Ge$_2$  & $P\overline{3}m1^{*}$, $Cmmm^{**}$\\
$I\overline{4}3d$$^{n,i,j}$ & 3.75 &   Li$_{15}$Ge$_4$  & $I\overline{4}3d$\\
$F\overline{4}3m$$^{o}$   & 4.20 & Li$_{17}$Ge$_4$  & $F\overline{4}3m$ \\
$F\overline{4}3m^{p,j,q}$     & 4.25 &Li$_{22}$Ge$_5$  &   $F\overline{4}3m^*$  \\
$P6_3/mmc$$^r$   & -- & $\alpha$Li & $P6_3/mmc$\\
\hline\hline
\end{tabular}
\begin{flushleft}
Sangster and Pelton's work was invaluable for an overview of the field. \cite{sangster:JPE:1997}\\
$^*$ First metastable above tie-line.\\
$^{**}$ Second metastable above tie-line.\\
$^{a}$ A.~W. Hull \cite{hull:PR:1922}.\\
$^b$  Very brief summaries are given by Gr\"uttner \etal \cite{gruttner:ACA:1981,gruttner:ACIE:1982}\\
$^c$ Kiefer and F\"assler. \cite{kiefer:SSS:2011}\\
 $^{d}$ E. Menges \etal\ \cite{menges:ZN:1969}\\
$^{e}$ J. Evers, \etal\ \cite{evers:ACIE:1987}\\
$^f$ Reported in abstract by Gr\"uttner \etal \cite{gruttner:ACA:1981}\\
$^g$ First found by Frank \etal \cite{frank:ZN:1975}; Nesper \etal\ \cite{nesper:JSSC:1986} suggested it is actually Li$_8$MgGe$_6$.\\
$ ^{h}$ V. Hopf \etal\, \cite{hopf:ZN:1970}\\
$^i$ Jain \etal \cite{jain:JPPC:2013}\\
 $^j$ Yoon \etal \cite{yoon:ESSL:2008}\\
$^l$ E.~M. Pell finds Li$_3$Ge$_1$.\cite{pell:JPCS:1957} See Sangster \etal\ $^a$ and a discussion therein. St. John \etal\ report that they have found the Li$_3$Ge$_1$ reported earlier as Li$_{16}$Ge$_{5}$. \cite{st.john:JES:1982}\\
 $^{m}$ V. Hopf \etal\, \cite{hopf:ZN:1972}\\
 $^{n}$ Gladyshevskii \etal \cite{gladyshevskii:SPC:1961} and Johnson \etal\ \cite{johnson:AC:1965}\\
 $^o$ Goward \etal \cite{goward:JAC:2001}\\
 $^{p}$ Gladyshevskii \etal\cite{gladyshevskii:SPC:1964}\\
 $^q$ Reported by Jain \etal \cite{jain:JPPC:2013} as  Li$_{21.1875}$Ge$_5$\\
$^r$ C.~.S. Barrett.\cite{barret:PR:1947}\\
\end{flushleft}
\end{table*}

In order of lithium content, the following Li-Ge phases have all been proposed: Li$_7$Ge$_{12}$, Li$_1$Ge$_1$,
Li$_{12}$Ge$_7$, Li$_{11}$Ge$_6$, Li$_9$Ge$_4$, Li$_7$Ge$_3$,
Li$_7$Ge$_2$, Li$_{15}$Ge$_{4}$, Li$_{17}$Ge$_4$ and Li$_{22}$Ge$_5$.
Below we compare in detail the known phases to the results of the DFT searches.

Li$_7$Ge$_{12}$ is the only reported phase with a ratio of Li/Ge less than 1.
It has two symmetries associated with it, originally $Pmn2_1$ \cite{gruttner:ACA:1981,gruttner:ACIE:1982}, which was later disputed,\cite{sangster:JPE:1997} and more recently, $P2/n$.\cite{kiefer:SSS:2011}
Four of its lithium lattice sites are 50\% occupied.
We model its structure in a periodic lattice using a simulation cell containing 28 Li and 48 Ge sites.
The fractionally occupied sites can be filled in a variety of ways: all
sites, giving rise to a crystal symmetry $P2/c$: one site ($P1$), four
different ways (all degenerate): two sites, 6 ways ($Pc$, $P2$ or
$P\overline{1}$) each symmetry being doubly degenerate): three sites ($P1$) four
ways (all degenerate): and by leaving all empty ($P2/c$) one way.
A convex hull of their single point energies shows that the ($Pc$) version is the most stable, hence we use this throughout the rest of the calculations.
Although not on the Li-Ge convex hull, see Fig.\,\ref{Fig:LiGe_hull}, this $Pc$ predicted phase is close above.

Li$_1$Ge$_1$ has an $I4_1/a$ \cite{menges:ZFNB:1969} ground state and a $I4_1/amd$ \cite{evers:AC:1987} high pressure form.
DFT predicts $I4_1/amd$ and a new layered $P4/mmm$ phase $\sim0.012$\,eV/f.u. and $\sim0.020$\,eV/f.u. above the $I4_1/a$ ground state respectively.
The $P4/mmm$ phase is isostructural with the Li-Si phase discussed above in Sec.\,\ref{Sect:Li1Si1}.
DFT predicts that the system undergoes a phase transition from the $I4_1/amd$ to the  $P4/mmm$ phase at 5\,GPa.

\begin{table*}
\caption{\label{table3} Low energy Li$_1$Ge$_1$ metastable phases, with formation energy, $E_f$ p.f.u relative to that of the energy of the ground state. The structures are isotypic of those found in Li$_1$Si$_1$, as shown in Table\,\ref{Table:Li1Si1} and Fig.\,\ref{Fig:LiSi_montage}. DFT predicts that the P4/mmm phase is the most stable above 5\,GPa.}
\begin{tabular}{ @{} c  c c c}
\hline\hline
E$_f$ &  Symmetry & Volume  & Description \\
(eV/f.u.) & & (\AA$^3$/f.u.)  & \\
\hline
0.00    &   $I4_1/a$  &  35.1   &   Li tetrahedra in a 3-fold coordinated Ge network\\
0.01   &   $I41/amd$ & 32.2 &  Known high pressure phase\\
0.02  & $P4/mmm$ & 31.9 & Flat Ge sheets 4 membered rings \\ 
0.03 & $R\overline{3}$ & 36.8 & Distorted Li octahedra 3-fold coordinated Ge network \\
0.04 & $P2/m$ & 32.1 & Isostructural with the corresponding Li$_1$Sn$_1$ phase\\
\hline\hline
\end{tabular}
\end{table*}

Gr\"uttner \etal\ mentioned a Li$_{12}$Ge$_7$ phase isotypic with the corresponding Li$_{12}$Si$_7$ phase \cite{gruttner:ACA:1981} in a very brief report, but did not present any further data to support its discovery. 
DFT also predicts a  Li$_{12}$Ge$_7$ $Pnma$ phase near the tie-line. 

Li$_{11}$Ge$_6$ was synthesized by Frank \etal \cite{frank:ZN:1975} with a molecular volume of 172.3\,cm$^3$mol$^{-1}$. 
Nesper \etal\ \cite{nesper:JSSC:1986} claim that the phase is actually Li$_8$MgGe$_6$, suggesting that since Li$_8$MgGe$_6$ has a molecular volume of 166.5\,cm$^3$mol$^{-1}$ it is unlikely that Li$_{11}$Ge$_6$ could have two more atoms per formula unit.\footnote{Confusingly Sangster and Pelton \cite{sangster:JPE:1997} report that Nesper \etal\ claim it is the Li$_8$MgSi$_6$ phase.}
DFT also finds a Li$_{11}$Ge$_6$ $Cmcm$ phase slightly above the tie-line with a volume of 286.26\,\AA$^3$ per f.u., which corresponds to a  molecular volume of 172.4\,cm$^3$mol$^{-1}$. 
Hence it seems entirely possible to us that Li$_{11}$Ge$_6$ $Cmcm$ was synthesized by Frank \emph{et al.} as initially proposed.

Li$_9$Ge$_4$ in the $Cmcm$ symmetry group, have been made electrochemically and from high temperature fusion, \cite{sangster:JPE:1997,jain:JPPC:2013,yoon:ESSL:2008} but all of our
calculations show it well above the tie line, favoring disproportion into a  $P32_12$ Li$_7$Ge$_3$ phase. 
Li$_7$Ge$_3$ with $P32_12$ symmetry was first mentioned by ref.\,\onlinecite{gruttner:ACA:1981} but no supporting information was given.  Jain \etal\ found an unknown phase that they suggested was Li$_7$Ge$_3$ fitting the diffraction data to $R\overline{3}m$ symmetry.\cite{jain:JPPC:2013}
Hence we suggest that Jain \etal\ synthesized either the Li$_5$Ge$_2$ or indeed Li$_8$Ge$_3$ phases, DFT predicting that both phases have the $R\overline{3}m$ symmetry.
Li$_5$Ge$_2$ is above the tie-line and Li$_{13}$Ge$_5$ and Li$_8$Ge$_3$ are all stable although to the best of our knowledge they have not previously been presented in the literature.
This may be due to thermal effects, see the discussion of similar arguments for Li$_5$Si$_2$ in Sec.\,\ref{Sect:Li5Si2}.
Li$_{16}$Ge$_5$ was predicted by St.~John, \etal\ \cite{st.john:JES:1982} during electrochemical studies.
They presented no crystal structure nor is there any prototype structure of \{Li/Na\}\{Si/Ge/Sn/Pb\} with this stoichiometry.
DFT predicts a Li$_{13}$Ge$_4$ $Pbam$ phase, isostructural with the Li$_{13}$Si$_4$ phase which is slightly above the tie-line and with a similar Li:Si ratio to Li$_{16}$Ge$_5$.
Recently preliminary results by H. Jung \etal\ \cite{Jung:Unp:2013} have produced electrochemically new phases in the Li$_{2.33}$Ge - Li$_{3.5}$Ge range whose X-ray pair distribution functions (PDF) match at least one of our predicted phases.
A fuller investigation will be presented later.

The Li$_7$Ge$_2$ phase with $Cmmm$ symmetry can be made electrochemically and by annealing from high temperature melt. \cite{sangster:JPE:1997,yoon:ESSL:2008} 
DFT-GGA predicts the $P\overline{3}m1$ phase above the tie-line and 0.08\,eV/f.u more stable than the  $Cmmm$ phase.
This discrepancy remains after using harder pseudopotentials and either the LDA exchange-correlation or the HSE06 hybrid functional. \cite{krukau:JCP:2006}
A fuller investigation into this will be presented elsewhere.

We find the well known Li$_{15}$Ge$_4$ stoichiometry $I\overline{4}3d$ phase stable.\cite{jain:JPPC:2013,yoon:ESSL:2008,gladyshevskii:SPC:1961,johnson:AC:1965}
The most lithiated phase has been a matter of debate in all Li-Group 4 compounds including germanium.
Its stoichiometry was reported as Li$_{22}$Ge$_5$ with $F23$ symmetry,\cite{gladyshevskii:SPC:1964} due to its similarity to Li$_{22}$Pb$_4$. \cite{zalkin:JPC:1958}
More recently, Goward \etal\ \cite{goward:JAC:2001} studied this family of structures and show that for the Ge, Sn and Pb compounds the correct stoichiometry is Li$_{17}$Ge$_4$ with $F\overline{4}3m$ symmetry.
$F\overline{4}3m$ symmetry Li$_{21}$Ge$_5$, Li$_{22}$Ge$_5$ and Li$_{17}$Ge$_4$ are found by DFT all  at local energy minima.
However, Li$_{17}$Ge$_4$ is on the tie-line.
Fassler \etal also predict a Li$_{4.10}$Ge phase analogous to the Li$_{4.11}$Si phase.
We use the same model structures as in the Li$_{4.11}$Si phase for our DFT calculations.
These are above the tie-line as is expected for a high-temperature phase.

The Li-Ge system has an analogous inclination to forming Ge-Ge dumbbells as in Li-Si, as discussed in Sec.\,\ref{dumbells}.

\begin{figure}
\includegraphics*[width=90mm]{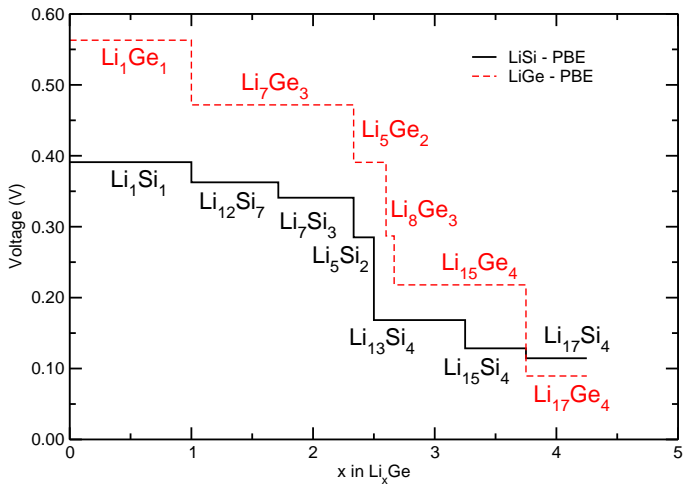}
\caption[]{(Color online) Potential-composition curves of stable structure found on the Li-Si and Li-Ge convex hull diagrams. The Li-Ge results are at a higher average voltage.}
\label{Fig:LiGe_voltages}
\end{figure}

\section{Discussion}
\label{Sect:Conclusions}

Crystal structures of the Li-Si and Li-Ge systems have been presented, found using both AIRSS searches and atomic species swapping of ICSD structures.  
Below we discuss the structures that are likely to be thermally accessible at room temperature, that is, those at a local minima on the DFT potential energy surface which reside on, or close to, the convex hull.
These structures serve as a model for the clustering and bonding behavior of electrochemically lithiated silicon and germanium.\cite{ogata:NC:2014}

The Li-Si system was used to validate our method: DFT finds all of the known phases as local energy minima including independently uncovering the Li$_{17}$Si$_4$ phase. 
For the Li-Ge system, DFT finds Li$_5$Ge$_2$, Li$_8$Ge$_3$, Li$_{13}$Ge$_5$ and Li$_{13}$Ge$_4$ locally stable and, to the best of our knowledge, these have not been presented in the literature before.
DFT predicts that Li$_{7}$Ge$_{12}$ and Li$_{11}$Ge$_6$ are local energy minima; the former having $Pc$ symmetry and the latter $Cmcm$.
It was reported that Li$_{11}$Ge$_6$ may be produced from a high-temperature melt \cite{frank:ZN:1975} but this has been disputed.\cite{nesper:JSSC:1986}
Also at local energy minima are the Li$_{12}$Ge$_7$ $Pnma$ and Li$_7$Ge$_3$ $P32_12$ phases, which were suggested by Gr\"{u}ttner \cite{gruttner:ACA:1981} but without presenting the crystal structure.
An unknown phase was found by heating ball milled Li-Ge, its XRD pattern fits an Li$_7$Ge$_3$ phase with $R\overline{3}m$ symmetry.
Since DFT and  Gr\"{u}ttner both predict Li$_7$Ge$_3$ has $P32_12$ symmetry, we propose that the unknown phase may be either the Li$_5$Ge$_2$ or Li$_8$Ge$_3$ phases which have a similar stoichiometry to Li$_7$Ge$_3$ and both of which DFT predicts to have $R\overline{3}m$ symmetry.

For the Li-Si and Li-Ge structures on the tie-lines, the average voltages were calculated relative to lithium metal.  
This included for the first time Li$_{17}S$i$_4$ and Li$_1$Si$_1$.
The average voltages are in good agreement with both previous calculations and experiment.
They are higher in Li-Ge than Li-Si, implying that germanium has a lower energy density than silicon. 
However the higher insertion voltage is safer during lithiation, reducing the chance of lithium plating which can result in dendrites short circuiting the cell.
Lithium in germanium also has higher diffusivity than in silicon.

Li$_1$Si$_1$ was previously only synthesisable at high pressure but has recently been synthesized by highly energetic ball milling, remotivating interest in the high-pressure phases.
AIRSS searches predict a selection of higher energy Li$_1$Si$_1$ and Li$_1$Ge$_1$ phases.
At lower pressures three-dimensional three-fold coordinated silicon/germanium networks were prevalent. 
However, at higher densities, both silicon and germanium exhibited a $P4/mmm$ structure comprising flat sheets of four-fold coordinated silicon and germanium atoms respectively.
These became the most stable phase of Li$_1$Si$_1$ and Li$_1$Ge$_1$ at 2.5\,GPa and 5\,GPa respectively.
Given the interest in silicene our layered compounds might provide an alternative route to layered silicon.

A LIB does not necessarily have time to equilibrate thermodynamically over large length scales. \cite{key:JACS:2009}
The ability to generate a wide range of locally-stable low-energy structures above the ground state allows us to visualize the types of clusters which form in the LIB during cycling.
Over a lithiation range of Li$_x$Si, $x=2.33 - 3.25$ we found that the structures present exhibited Si-Si dumbbells. 
At higher lithiation all of the silicon dumbbells break up and the crystalline Li$_{15}$Si$_4$ phase forms.
Since these dumbbells were seen in both ground state and metastable phases is seems likely that they will exist in LIB anodes, probably in a lower symmetry solid solution.
Furthermore we find the analogous dumbbell containing structures in the Li-Ge system.

Above we have demonstrated that the combination of both atomic species swapping the ICSD phases and AIRSS is a powerful tool for predicting the crystal structures of LIB electrode materials.
A refinement to the method combines these two techniques by using results of the AIRSS searches as inputs to the species swapping technique.
For example, the low-energy structures found by AIRSS in Li$_1$Si$_1$ were re-optimized as candidate Li$_1$Ge$_1$ phases in the Li-Ge system.

Our method has only provided results of the stable and metastable structures at 0\,K, of course, the effect of temperature could be included \emph{post hoc} using phonon calculations within the harmonic approximation and beyond.
Our method serves as a crucial first step in \emph{ab initio} materials discovery and design.

\begin{acknowledgements}
The authors would like to thank Edgar Engel for useful discussions and Hyeyoung Jung, Yan-Yan Hu and Phoebe K. Allan for sharing their preliminary results and useful discussions.
AJM acknowledges the support from the Winton Programme for the Physics of Sustainability.
This work was supported by the Engineering and Physical Sciences Research Council (EPSRC) of the U.K. 
Computational resources were provided by the University College London Research Computing service and the Cambridge High Performance Computing service.
\end{acknowledgements}

\end{document}